\documentclass[aps,pra,twocolumn,superscriptaddress,showpacs]{revtex4-1}
\usepackage{amsmath}
\usepackage{amsmath}
\usepackage{amsfonts}
\usepackage{graphicx,color}
\usepackage{color}
\usepackage[bottom]{footmisc}
\usepackage{epstopdf}
\usepackage{epsfig}
\usepackage{graphics}
\usepackage[caption=false]{subfig}
\usepackage{url}
\usepackage{placeins}
\makeatletter

\newcommand{\tr}{\text{Tr}}

\newcommand{\Rmnum}[1]{\expandafter\@slowromancap\romannumeral #1@}
\makeatother

\begin{document}
\title{Engineering single-phonon number states of a mechanical oscillator via photon subtraction}
\author{M. Miskeen Khan}
\email[Electronic Address:\,]{mmiskeenk16@hotmail.com}
\affiliation{Department of Electronics, Quaid-i-Azam University, 45320 Islamabad, Pakistan}
\author{M. Javed Akram}
\email[Electronic Address:\,]{mjakram@qau.edu.pk}
\affiliation{Department of Electronics, Quaid-i-Azam University, 45320 Islamabad, Pakistan}
\author{M. Paternostro}
\email[Electronic Address:\,]{m.paternostro@qub.ac.uk}
\affiliation{Centre for Theoretical Atomic, Molecular and Optical
Physics, School of Mathematics and Physics, Queen's University,
Belfast BT7 1NN, United Kingdom}
\author{F. Saif}
\email[Electronic Address:\,]{farhan.saif@qau.edu.pk}
\affiliation{Department of Electronics, Quaid-i-Azam University, 45320 Islamabad, Pakistan}
\affiliation{Department of Physics, Quaid-I-Azam University, 3rd  Avenue, Islamabad}

\date{\today}

\begin{abstract}
We introduce an optomechanical scheme for the probabilistic preparation of single-phonon Fock states of mechanical modes based on photo-subtraction. The quality of the produced mechanical state is confirmed by a number of indicators, including phonon statistics and conditional fidelity. We assess the detrimental effect of parameters such as the temperature of the mechanical system and address the feasibility of the scheme with state-of-the-art technology. 
\end{abstract}

\maketitle


Recent developments in quantum optomechanics~\cite{aspelmeyer2014cavity} have shown the promises for quantum state engineering held by various experimental platforms. Squeezing of the quantum noise of a micromechanical resonator has been recently demonstrated in at least two experimental settings~\cite{pikkalainen,wollman}, while the first steps towards optomechanical entanglement have been reported in some noticeable experiments~\cite{palomaki, riedinger}.

Although the investigations performed so far have focused on Gaussian operations~\cite{schmidt} and states, including the engineering of universal resources for quantum computation~\cite{houhou}, similar attention has been paid to the preparation of non Gaussian states~\cite{zurek2003decoherence,khalili2010preparing,akram2010single,paternostro2011engineering,PhysRevA.83.013803,tan2013deterministic,asjad2014reservoir,PhysRevA.91.013842,akram2013entangled,PhysRevA.89.053829}. In this respect, the preparation of phononic number states is  a particularly important goal in light of the possibility to use optomechanical devices as memories and on-demand single-photon sources~\cite{galland2014heralded}. 
Such states have  been realized experimentally by coupling the mechanical mode to a superconducting qubit~\cite{o2010quantum}.

In this paper, we propose a novel scheme for the preparation of single-phonon number states (SPNS) that combines the features offered by the linearized optomechanical interaction and the potential for effective non-linear effects made available by photon subtraction. In Ref.~\cite{paternostro2011engineering}, one of us has demonstrated the effectiveness of photon subtraction for the preparation of non-classical states of mechanical modes: The non-classical correlations established between mechanical and optical oscillators in an optomechanical cavity, complemented by the subtraction of a single photon from the optical field, are sufficient to prepare the mechanical mode in a highly non-classical (non-Gaussian) state. Here, we extend such scheme showing the possibility to engineer a state that is very close to an ideal SPNS in both a dynamical way and at the steady-state of the optomechanical evolution. We characterize the quality of the resource thus achieved using both state fidelity and the phonon-number statistics.

The remainder of this paper is organized as follows: In Sec.~\ref{sec:II}, we introduce the system under scrutiny and its equations of motion. In Sec.~\ref{sec:IV}, we provide the analytical form of the conditional mechanical state achieved after a single-photon subtraction on the optical field. Sec.~\ref{sec:V} is devoted to the characterisation of such conditional state. Sec.~\ref{sec:new} addresses the steady-state version of the scheme put forward here, while Sec.~\ref{S6}, summarises our work and pinpoints a few open questions that are left for further investigation.

 \section{The Model and its dynamics}\label{sec:II}
For the sake of definiteness, we consider a single-mode Fabry-P\'erot cavity endowed with a movable mirror, although our considerations are valid for any other optomechanical system working in the regime where the position of a mechanical oscillator is linearly coupled to the intensity of the field accommodated in the cavity. The field mode has frequency $\omega_{c}$, and the cavity decay rate is $\kappa$. The mechanical mode oscillates at frequency $\omega_{m}$ and is affected by a local thermal environment (at temperature $T$) with which it exchanges excitations at a rate $\gamma_{m}$~\cite{aspelmeyer2013cavity}. The cavity is pumped by an external laser field at frequency $\omega_0$. 
In a frame rotating at such frequency, the Hamiltonian of the system reads
\begin{equation}\label{eq:Hamiltonian}
 \hat{H}_{s}=\hbar\Delta_{0}\hat{a}^{\dagger}\hat{a}+\frac{\hbar\omega_{m}}{2}(\hat{q}^2+\hat{p}^2)-\hbar G_{0}\hat{q}\hat{a}^{\dagger}\hat{a}+i\hbar E(\hat{a}^{\dagger}-\hat{a}), 
 \end{equation}
 \begin{figure}[t]\label{fig1}
\begin{center}
\includegraphics[width=0.4\textwidth]{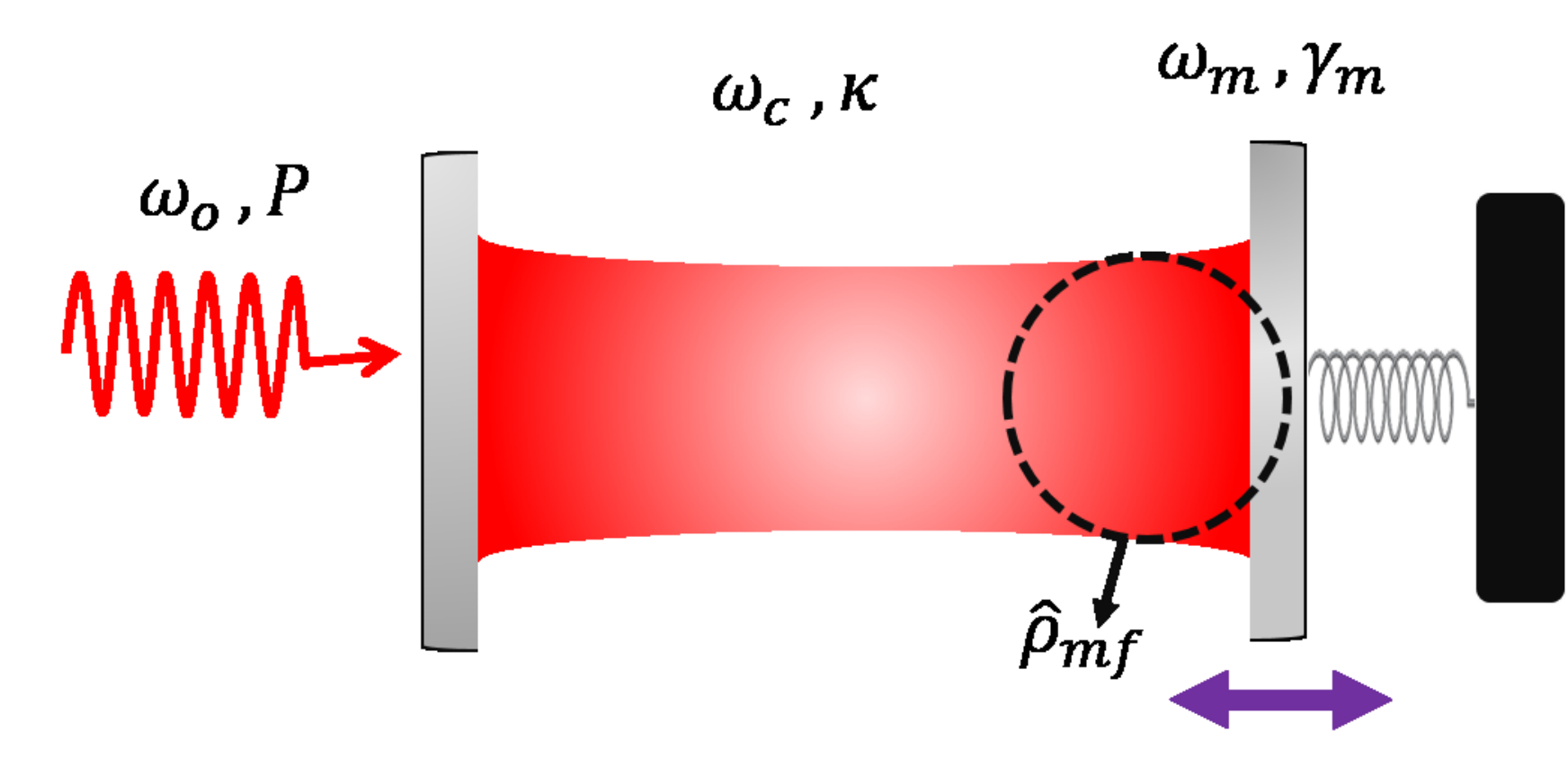}
 \caption{(Color online) Paradigm of an optomechanical system. An optical mode of frequency $\omega_{c}$ is confined within a cavity with decay rate $\kappa$. The mechanical mode associated with a movable mirror has frequency $\omega_{m}$ and is coupled to its local environment with rate $\gamma_m$. The cavity mode is excited with an input laser of frequency $\omega_{0}$ and power $P$. The joint state of both modes is labelled as $\hat{\rho}_{mf}$.}
 \label{fig:cavity}
\end{center}
\end{figure} 
where $\Delta_{0}=\omega_{c}-\omega_{0}$ is the cavity-pump detuning, $\hat{a}(\hat{a}^{\dagger})$ is the annihilation (creation) operator of the cavity field, $\hat{q}$ and $\hat{p}$ are the dimensionless position and momentum operators for the mechanical oscillator, and we have introduced the single-photon optomechanical coupling strength $G_{0}=\frac{\omega_{c}}{L}\sqrt{\frac{\hbar}{m\omega_{m}}}$ with $m$ the effective mass of the oscillator and $L$ the length of the cavity. Finally the last term in Eq.~\eqref{eq:Hamiltonian} describes the laser interaction with the cavity mode, which occurs at a rate $E=\sqrt{\frac{2P\kappa}{\hbar\omega_{0}}}$ with $P$ the power of the driving field.

Besides the unitary dynamics generated by Eq.~\eqref{eq:Hamiltonian}, we shall also consider the non-unitary one due to the coupling of the cavity with its electromagnetic environment and of the mechanical mode with the thermal background of phononic modes provided by the support upon which it is fabricated. 
For a large enough input power, the dynamics of the system can be split in the evolution of the mean fields of the system and that of the corresponding fluctuations. In the limit of large mechanical quality factor, the latter evolve according to the equation
\begin{equation}\label{eq:linerset}
\partial_t\hat{\bf u}(t)={\bf k}\,\hat{\bf u}(t)+\hat{\bf n}(t),
\end{equation}
where $\hat {\bf u}(t)=(\hat{\delta q}\,\hat{\delta p}\,\hat{\delta X}\,\hat{\delta Y})^{T}$, and $\hat{\delta O}$ ($O=q,p,X,Y$) is the fluctuation associated with operator $\hat O$ [we have introduced the field quadratures $\hat X=(\hat a+\hat a^\dag)/\sqrt{2}$ and $\hat Y=i(\hat a^\dag-\hat a)/\sqrt2$]. Moreover, $\hat{\bf n}(t)$ 
accounts for the input noise to the system, and $\textbf{k}$ is the system kernel matrix~\cite{vitali2007optomechanical}.
 The formal solution of Eq.~\eqref{eq:linerset} is given by
\begin{equation}\label{eq:linersetsol}
\hat{\bf u}(t)=\textbf{M}(t)\hat{\bf u}(0)+\int_{0}^{t}d\tau \textbf{M}(\tau)\hat{\bf n}(t-\tau), 
\end{equation}
where $\textbf{M}(t)=e^{\mathbf{k}t}$. Stability of such solution is guaranteed by meeting specific conditions on the parameters of the system~\cite{paternostro2006reconstructing}, which we assume to be the case throughout the remainder of our analysis (the numerical simulations reported later on are all well within such stability domain). 
We now introduce the covariance matrix $\textbf{v}(t)$ with elements $\text{v}_{ij}(t)=\langle u_{i}(t)u_{j}(t)+u_{j}(t)u_{i}(t)\rangle/2 $ ($i,j={1,...,4}$), which fully describe the Gaussian state of the system at hand. 
With this notation, we can recast Eq.~\eqref{eq:linerset} as
~\cite{rogers2014hybrid,ferreira2009quantum}
\begin{equation}\label{eq:eomcovar}
\partial_t{\textbf{v}}(t)=\textbf{k}\textbf{v}(t)+\textbf{v}(t)\textbf{k}^{T}+\textbf{D},
\end{equation}
where $\textbf{D}=$Diag$[0,\gamma_{m}(2\overline{n}+1),\kappa,\kappa]$ is the system diffusion matrix that encompasses the statistical properties of the noise affecting the optomechanical system. 

\section{Conditional state of the mechanical mode} \label{sec:IV}
As discussed earlier, the main goal of the scheme is to subtract a single photon from the field reflected by the cavity end-mirror at a given instant of the evolution, and analyze the features of the conditional mechanical state. 

The joint optomechanical state $\hat{\rho}_{mf}$ can be written as
\begin{equation}
\hat{\rho}_{mf}=\frac{1}{\pi^{2}}\int d^{2}\lambda d^{2}\eta~C_{mf}(\lambda,\eta,t)\hat{D}^{\dagger}_{m}(\eta)\otimes\hat{D}_{f}^{\dagger}(\lambda),
\end{equation}
where $C_{mf}(\lambda,\eta,t)$ is the Weyl characteristic function of such joint state, $\hat{D}^{}_{m}(\eta)$ [$\hat{D}_{f}^{}(\lambda)$] is the displacement operator for the mechanical mode [cavity field], and $\eta=\eta_r+i\eta_i$ [$\lambda=\lambda_r+i\lambda_i$] is the corresponding phase space variable. As the overall state of the system is Gaussian, we have $C_{mf}(\lambda,\eta,t)= \exp[-\frac{1}{2}\textbf{x}\textbf{v}(t)\textbf{x}^{T}]$ with ${\bf x}=(\eta_r\,\eta_i\,\lambda_r\,\lambda_i)^T$.

We now assume that, at a given time of the joint evolution of the system, the cavity field mode is subjected to a single-photon subtraction process. This is formally implemented by the application of the annihilation operator $\hat a$ to the state of the cavity field. As we are only interested in the properties of the mechanical mode, we discard the optical state, finding the conditional mechanical density matrix 
\begin{equation}\label{conditionalsta}
\hat{\rho}_{m}=\frac{{\cal N}}{\pi^{2}}\int d^{2}\lambda d^{2}\eta~C_{mf}(\lambda,\eta)\hat{D}^{\dagger}_{m}(\eta)\tr_f[\hat{a}\hat{D}_{f}^{\dagger}(\lambda)\hat{a}^{\dagger}].
\end{equation}
Here, ${\cal N}$ is the normalization constant. We can further elaborate this expression by considering that 
\begin{equation}\label{tracecal}
\begin{aligned}
\tr[\hat{a}\hat{D}_{f}^{\dagger}(\eta)\hat{a}^{\dagger}]&=\frac{1}{\pi}\int d^{2}\alpha(\vert\alpha\vert^{2}-\vert\lambda\vert^{2}+\lambda^{\ast}\alpha-\alpha^{\ast}\lambda+1)\\
&\times \exp[-\frac{1}{2}\vert\lambda\vert^{2}+\lambda^{\ast}\alpha-\alpha^{\ast}\lambda].
\end{aligned}
\end{equation}
After some tedious but otherwise straightforward manipulations, the conditional density operator for the mechanical mode is found to be
\begin{equation}\label{conditionalstate}
\hat{\rho}_{m}=\frac{{\cal N}}{4\pi}\int d^{2}\eta \hat{D}^{\dagger}_{m}(\eta)g_{2}(\gamma)e^{g_{1}(\eta)},
\end{equation}
where we have introduced the functions 
\begin{equation}
\begin{aligned}
g_{1}(\eta)=&-\frac{1}{2} m_{22} \eta _i^2-\frac{1}{2} \eta _r \left(m_{12} \eta _i+m_{21} \eta _i+m_{11} \eta _r\right),\\
g_{2}(\eta)=&-[(c_{21}^2+c_{22}^2) \eta _i^2+(c_{11}^2+c_{12}^2) \eta _r^2\\
&+2(c_{11} c_{21} + c_{12} c_{22}) \eta _i \eta _r]+f_{11}+f_{22}-2.
\end{aligned}
\end{equation} 
Here $m_{ij}$ and $f_{ij}$ are the entries of the local covariance matrices of the mechanical and optical mode respectively, while the elements $c_{ij}$ encompass the cross-correlation between the two modes at hand. 
From Eq.~\eqref{conditionalstate} it is straightforward to evaluate the Wigner function of the conditional mechanical state as
\begin{equation}\label{wigner}
W(\delta_{r},\delta_{i},t)=A^t_{0}(A^t_1+B^t_{rr}\delta^2_r+B^t_{ri}\delta_r\delta_i+B^t_{ii}\delta^2_i)e^{C^t},
\end{equation}
where the time-dependent functions $A^t_{0,1}$, $B^t_{jk}~(j,k=i,r)$, and $C^t$ depend on the covariance matrix elements and are explicitly given in the Appendix. Clearly, Eq.~\eqref{wigner} displays the non-Gaussian character of the conditional mechanical state. Depending on the value taken by the polynomial, $W(\delta_{r},\delta_{i},t)$ can take negative values, thus signalling non-classicality~\cite{hudson1974wigner}. However, the time dependence of the functions in $W(\delta_{r},\delta_{i},t)$ makes the properties of the conditional mechanical state very sensitive to the exact time at which the photon subtraction is performed. In the next Section, we show the existence of a set of parameters and an instant of time at which the conditional mechanical state becomes very close to a SPNS.

\section{Results and Discussions}\label{sec:V}

\begin{figure}[t]
\begin{center}
\includegraphics[width=0.5\textwidth]{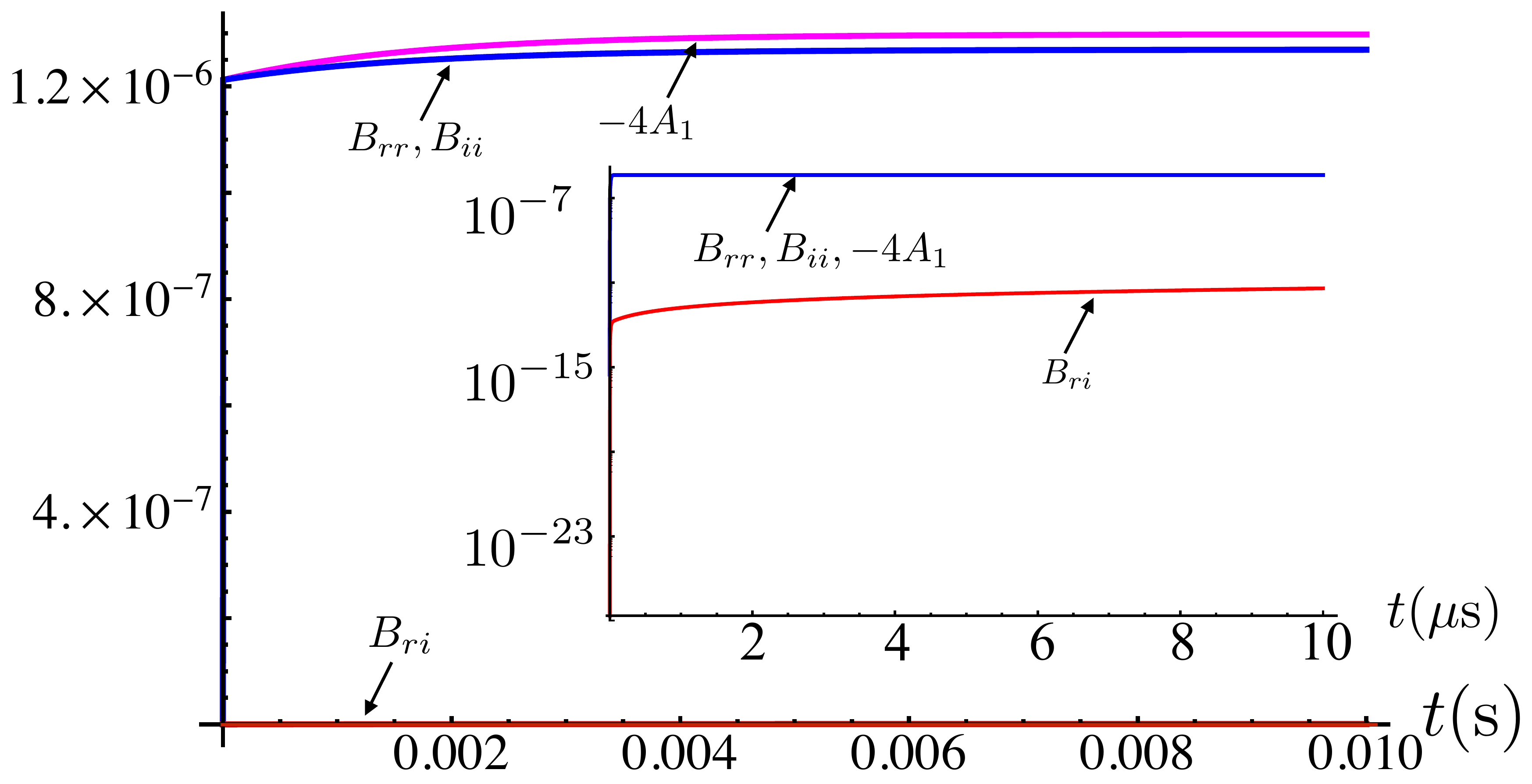}
 \caption{(Color online) Time dependence of the coefficients entering the polynomial in Eq.~\eqref{wigner} for $L=1$ mm, $\lambda=1064$ nm, $\omega_{m}/2\pi=1$ GHz, $P=5$ mW, $m=5$ ng, $\kappa/2\pi\simeq90$ MHz, $T=1$ mK, $\omega_{c}=\omega_{0}$, and $\gamma_{m}/2\pi=100$ Hz. We take $\Delta/\omega_{m}=-1$. The inset shows the same plot displayed in the main panel, but for a much shorter temporal window and in logarithmic scale. }
 \label{coefficients}
\end{center}
\end{figure}

We focus on parameters that are achievable experimentally~\cite{chan2011laser, chan2012optimized,galland2014heralded} (cf. Fig.~\ref{coefficients}) 
and consider high-frequency mechanical oscillators ($\sim 1$GHz, in line with proposals for the preparation of single-phonon states reported in Ref.~\cite{galland2014heralded}, and compatible with   photonic crystal nano-beam resonators~\cite{chan2012optimized}), operating at $1$mK (a temperature that, albeit beyond those achievable by means of standard dilution refrigerators,  can be reached by employing nuclear demagnetization refrigerators \cite{1742-6596-400-5-052024,ANDP:ANDP201400107}). We work in the blue detuned regime $\Delta_0<0$. Albeit this choice typically entails extra heating of the mechanical mode, leading the system to instability~\cite{hammerer2014nonclassical}, the set of parameters chosen for our numerical simulations guarantees stability of the system. Finally, we achieve weak-coupling and sideband-resolved conditions (i.e. $G/\kappa\ll1$ and $\kappa<\omega_{m}$ with $G=\sqrt{2}G_{0}\alpha_{s}$ the effective optomechanical coupling strength and $\alpha_{s}=\vert E\vert/\sqrt{\kappa^{2}+\Delta_0^{2}}$ the mean value of the cavity field).

\begin{figure}[b]
\begin{center}
\includegraphics[width=0.4\textwidth]{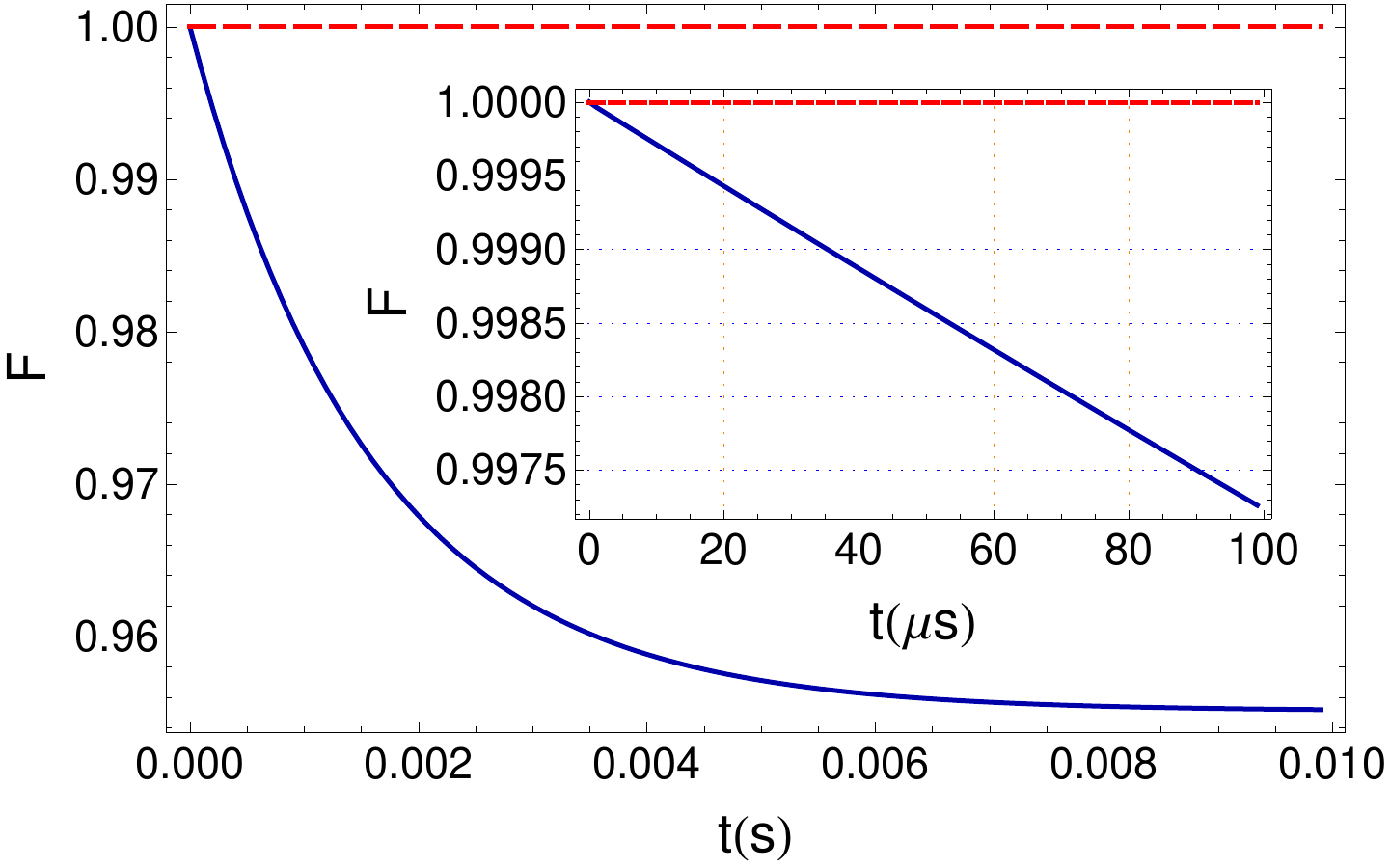}
 \caption{(Color online) State fidelity between the conditional mechanical state and a SPNS plotted against the time at which the photon subtraction is performed. We have $F\simeq1$ at short times, while a $4\%$ deviation from unity is seen for photon subtraction made at the steady state. All the parameters are the same as in Fig.~\ref{coefficients}.}
 \label{fig:fedility1}
\end{center}
\end{figure}

\begin{figure}[t]
\begin{center}
{\bf (a)}
\includegraphics[width=0.5\textwidth]{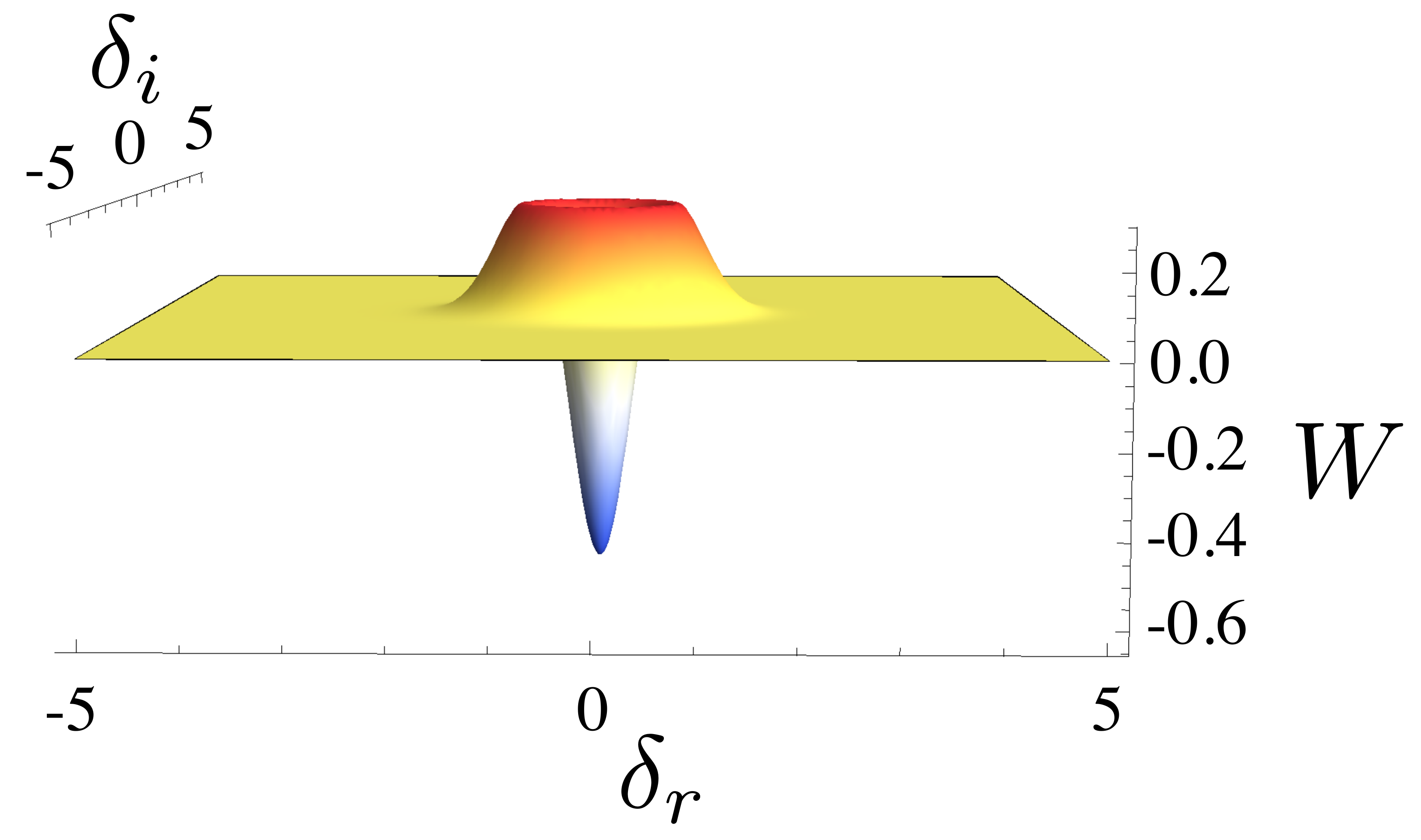}\\
{\bf (b)}
\includegraphics[width=0.5\textwidth]{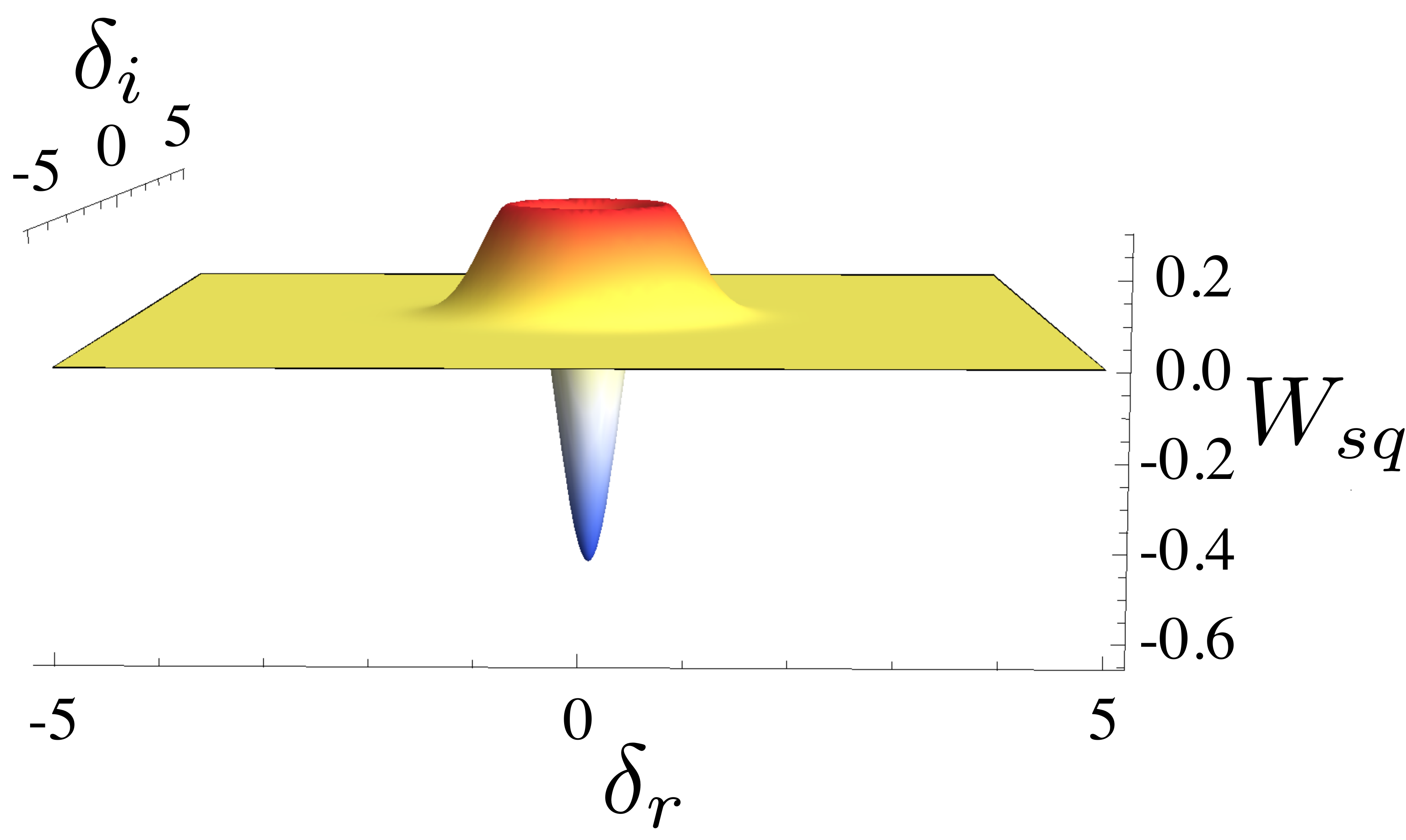}
 \caption{(Color online) Wigner function of the conditional mechanical state prepared through the proposed scheme and for the parameters used in Fig.~\ref{coefficients}. A single-photon subtraction from the field mode has been performed at the interaction time $\tau=9\mu$s. {\bf (b)} Wigner function of a SPNS.}
 \label{fig:wignercond}
\end{center}
\end{figure}

The goal of our investigation is to achieve a SPNS, whose Wigner function reads  
\begin{equation}\label{idealfock}
W_{sq}(\delta_r,\delta_i)=\frac{2  (4 \delta _i^2+4 \delta _r^2-1)}{\pi}e^{-2( \delta _i^2+ \delta _r^2)}.
\end{equation}
Therefore, in order to achieve such a target state, we should ensure that, at some time $\tau$, we have 
\begin{equation}
\label{conditions}
\begin{aligned}
B^\tau_{rr}\simeq B^\tau_{ii}\simeq-4A_1\quad\text{and}\quad B^\tau_{ri}\simeq0
\end{aligned}
\end{equation}
with the product  $A_{0}^\tau A_{1}^\tau$ providing the correct normalization. This is indeed the case for suitable choices of $\tau$: as shown in Fig.~\ref{coefficients}, while at the steady state and for the chosen set of parameters, the individual functions entering the polynomial in $W(\delta_r,\delta_i,t)$ do not satisfy the conditions stated in Eq.~\eqref{conditions} (albeit the discrepancies are small), an excellent agreement with the {\it desiderata} is instead achieved for $\tau$ up to $10\mu$s. We have also checked that, within such timeframe, we have $C^\tau\simeq-2(\delta^2_r+\delta^2_i)$, thus making the Wigner function of the conditional state close to the target one. Such considerations are made fully quantitative in Figs.~\ref{fig:fedility1} and \ref{fig:wignercond}.

The former presents the values taken by the state fidelity 
\begin{equation}
F=\pi\int d^{2}\delta~W(\delta_r,\delta_i,\tau)W_{sq}(\delta_r,\delta_i)
\end{equation}
against the time of photon subtraction. Clearly $F$ approaches unity for subtractions performed at short times. We find a numerical optimum ($F\sim0.99974$) at $\tau\simeq9\mu$s, a value that is retained in the remainder of our analysis. 
The latter shows the shape taken by the Wigner function of the conditional mechanical state for a photon subtraction taking place at such optimal time, and compares it to $W_{sq}(\delta_r,\delta_i)$, demonstrating that the two quasi probability distributions share the same features of rotational symmetry around the origin of the phase space, and the same amplitude of the negative peak at the origin. 

Let us now provide a physical intuition for the result that we have obtained. The blue-detuned regime $\Delta=-\omega_{m}$ that we have chosen corresponds to an effective optomechanical interaction dominated by a two-mode squeezing process~\cite{paternostro2008mechanism,genes2008robust}
\begin{equation}
\label{effectiveH}
\hat H_{mf,eff}\propto\hat{\delta q}\hat{\delta X}-\hat{\delta p}\hat{\delta Y}.
\end{equation}
In fact, by writing explicitly Eq.~\eqref{eq:linerset} in the blue detuned regime with $\Delta=-\omega_{m}$, it is straightforward to show that the mechanical oscillator and cavity field are coupled, in general, by a resonant process that simultaneously creates excitations in the mechanical and optical oscillator, and an off-resonant one that transfers excitations from one oscillator to the other. More explicitly, by calling $\hat{\delta b}$ ($\hat{\delta b}^\dag$) the annihilation (creation) operator of the mechanical oscillator in the interaction picture, the two processes above are linked to effective processes of the form 
$\hat{\delta b}^\dag\hat{\delta a}^\dag+\hat{\delta b}\hat{\delta a}$ (for the resonant mechanism) and $(\hat{\delta b}^\dag\hat{\delta a}e^{2i\omega_m t}+\hat{\delta b}\hat{\delta a}^\dag e^{-2i\omega_m t})$ (for the non-resonant one). By invoking the rotating-wave approximation, the second process is suppressed in favour of the first, which can then be recast as in Eq.~\eqref{effectiveH}~\cite{paternostro2008mechanism,genes2008robust}.

At short interaction times, when any environmental effect can be neglected, should the state of the mechanical system be close to the ground state, such interaction would result in a two-mode squeezed vacuum state of the fluctuations of the system. Let us then depart, momentarily, from the actual state of the system at hand and concentrate on the effect that uni-lateral excitation-subtraction has on the two-mode squeezed vacuum state of two bosonic modes, dubbed here as $1$ and $2$. It is straightforward to check that the conditional state of mode $1$ when $2$ is subjected to the subtraction of a single excitation reads
\begin{equation}
\rho_{1}=\sum^\infty_{n=0}P(n,s)|{n}\rangle\langle{n}|_1,
\end{equation}
where $s$ is the degree of two-mode squeezing of the unconditional state and  $P(n,s)=\frac{n(\tanh s)^{2n}}{[\cosh(s)\sinh(s)]^2}$ is the probability that state $|n\rangle$ is occupied. 
At small values of $s$, the probability distribution is sharply peaked around $n=1$, showing little contribution to the conditional state provided by highly excited number states. This picture breaks down as $s$ grows, given that more number states enter into the original two-mode squeezed vacuum state. Indeed, at low values of squeezing, a two-mode squeezed vacuum state is well approximated as $|\psi\rangle_{12}\propto|00\rangle_{12}+s|11\rangle_{12}$. The action of the subtraction operation on mode $2$ is thus equivalent to the heralded preparation of mode $1$ into a single-excitation state.
\begin{figure}[b]
\begin{center}
{\bf (a)}\\
\includegraphics[width=0.4\textwidth]{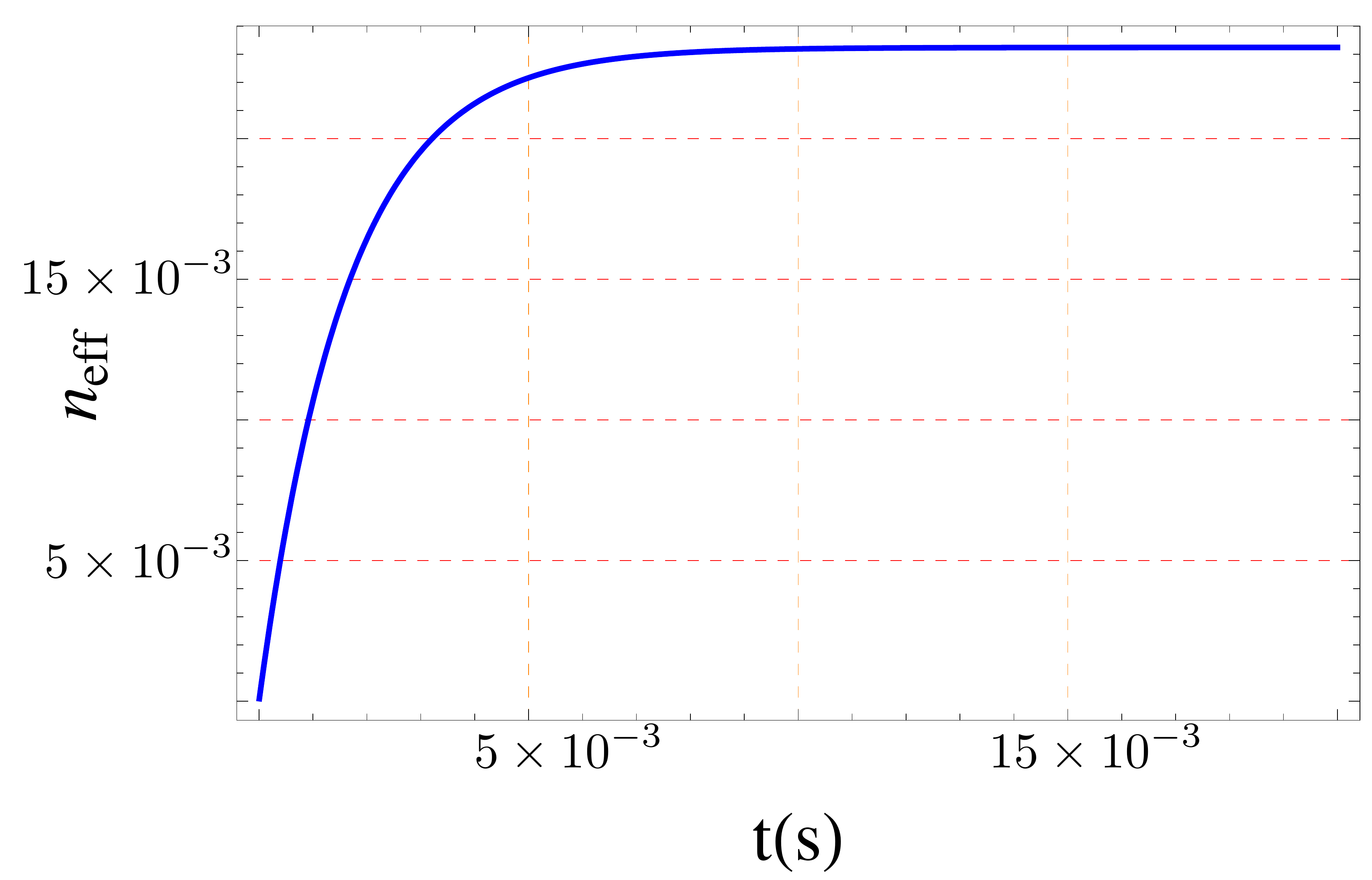}\\
{\bf (b)}\\
\includegraphics[width=0.4\textwidth]{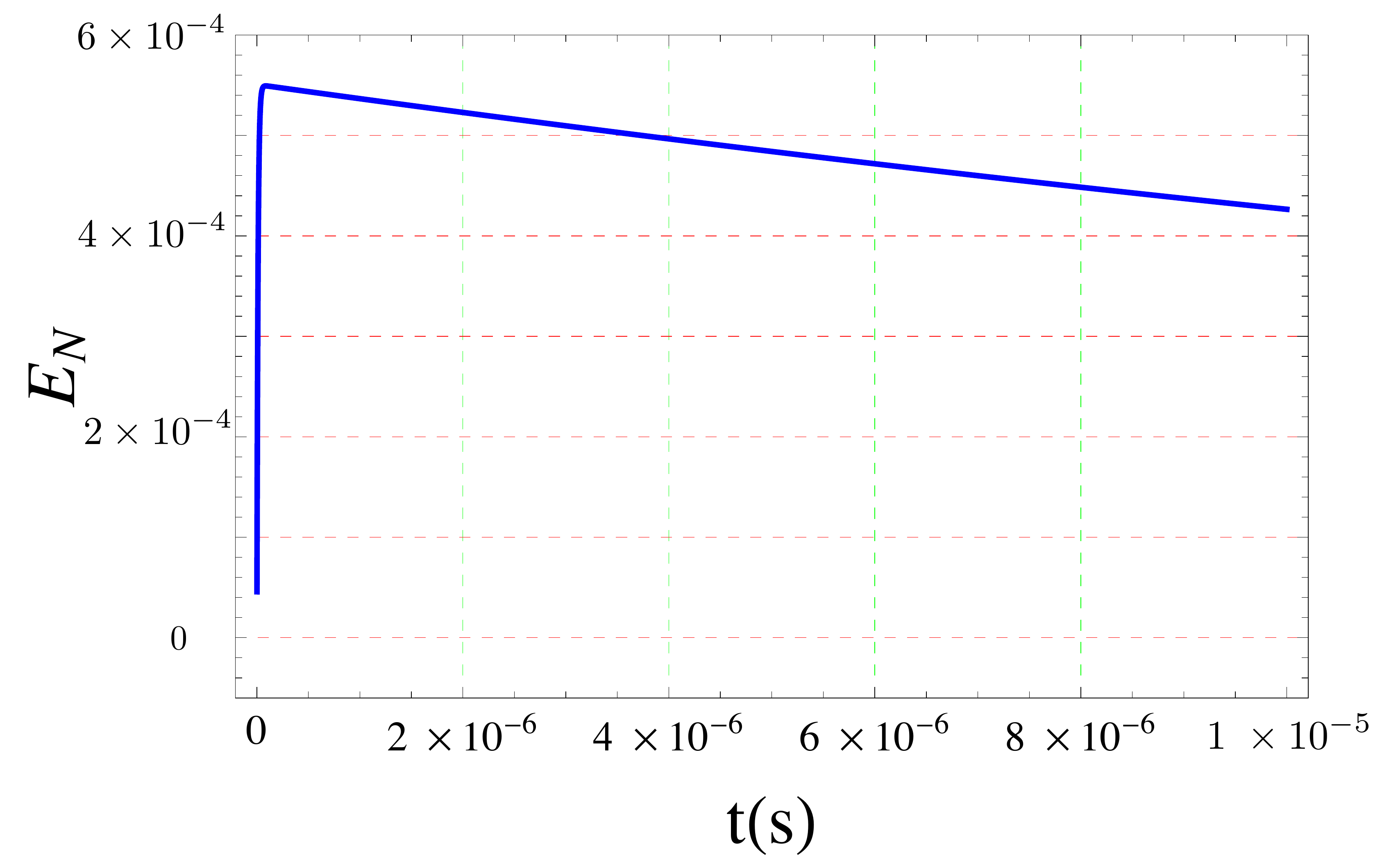}
 \caption{(Color online) {\bf (a)} We show the effective phonon number against the interaction time. {\bf (b)} Logarithmic negativity of the optomechanical state against the interaction time. In both panels, all the parameters are the same as in Fig. \ref{coefficients}.}
 \label{fig:effphonon}
\end{center}
\end{figure} 

In order to confirm the validity of such interpretation for the mechanism behind the success of our scheme, we should validate the assumptions that the mechanical mode is in a low-occupation state and that the degree of two-mode squeezing of the unconditional state is indeed very low. Both these assumptions are indeed verified by the analysis reported in Fig.~\ref{fig:effphonon}, where we show  the mean phonon number $n^t_\text{eff}=(m_{11}+m_{22}-1)/2$ in the mechanical state and the degree of optomechanical entanglement $E_N$ (quantified by the logarithmic negativity), both against the interaction time. Not only the mechanical system is always very close to its ground state, but also the degree of entanglement is always kept at very low levels. This is indirect confirmation of the small degree of equivalent two-mode squeezing of the optomechanical state. In fact, for $s\ll1$, the logarithmic negativety of a two-mode squeezed vacuum state is a linear function of $s$. Incidentally, Fig.~\ref{fig:effphonon} {\bf (a)} demonstrates that, although the blue-detuning regime chosen here is indeed responsible for the heating of the mechanical system, the corresponding phononic mean occupation number remains at low values all the way down to steady-state conditions, thus validating our chosen parameter regime. Finally, the need for a short interaction time to optimize the performance of the proposed protocol is due to the fact that, as time grows, the effects of the environments to which the system is exposed start becoming relevant and the simplified Hamiltonian picture of two-mode squeezing breaks down, leading the system to a steady state that is significantly different from a two-mode squeezed vacuum. Such considerations are supported by the analysis of the entanglement set between the mechanical oscillator and the cavity field. In Fig.~\ref{fig:effphonon} {\bf (b)} we show the temporal behavior of the logarithmic negativity $E_N$~\cite{horo}, which is an entanglement monotone perfectly suited to characterise the entanglement of two-mode Gaussian states such as the optomechanical one before the subtraction process. Some entanglement builds at very short times ($<1~\mu$s), and then decays due to the open-system dynamics undergone by the system and the subsequent deviation of the effective dynamics from the simple two-mode squeezing process in Eq.~\eqref{effectiveH}.

We now address the robustness of the state-engineering mechanism illustrated here to the effects caused by a larger temperature of the mechanical system. 
\begin{figure}[t]
\begin{center}
\includegraphics[width=0.85\linewidth]{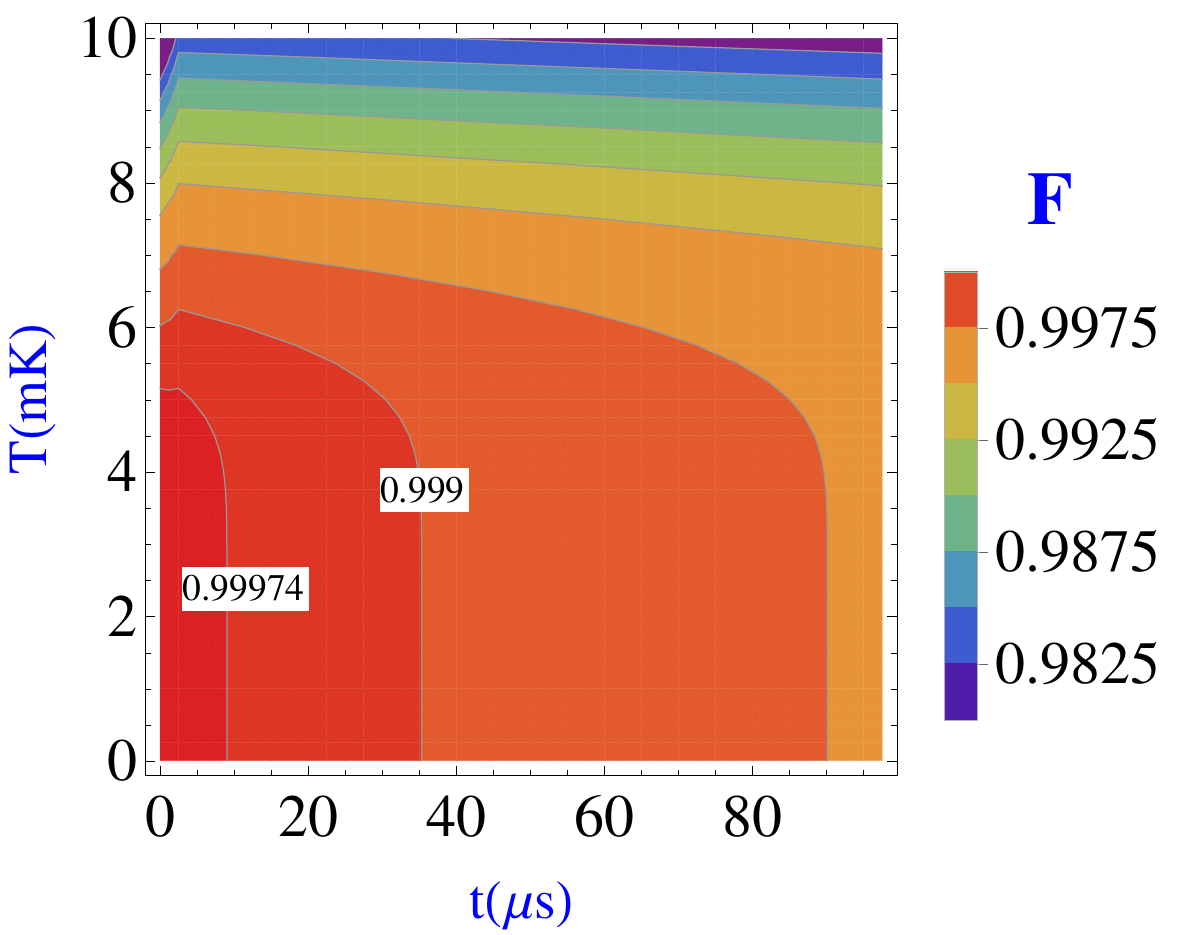}
\caption{(Color online) State fidelity against the initial temperature of the mechanical system and the (dimensionless) interaction time. The fidelity decreases as the photon subtraction is performed on a system with large initial temperature. All the parameters are the same as in Fig. \ref{coefficients}.}
\label{fig:fedilitya}
\end{center}
\end{figure} 
\begin{figure}[t]
\begin{center}  
\includegraphics[width=0.5\textwidth]{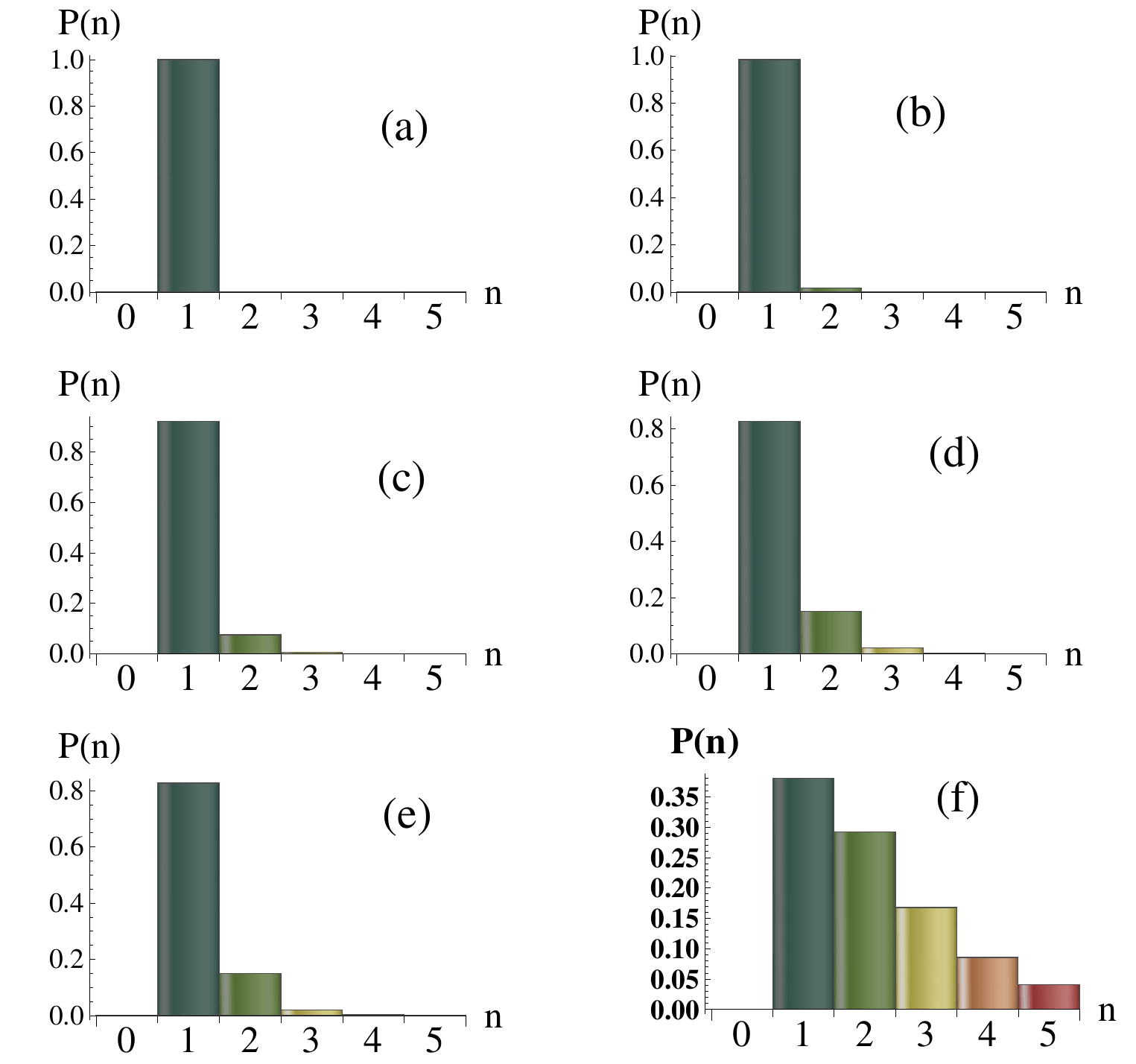}
\caption{(Color online) Phonon-number distribution for the different values of the temperature. We have taken (a) $T=5$mK, (b) $T=10$mK, (c) $T=15$mK, (d) $20$mK, (e) $T=25$mK, and (f) $T=50$mK. All the other parameters are the same as in Fig.~\ref{coefficients}.}
 \label{fig:phostattemp}
\end{center}
\end{figure} 
 In Fig.~\ref{fig:fedilitya} we show the degradation of the state fidelity as the temperature of the initial state of the mechanical system grows. We specifically highlight the contours for $F=0.99974$ and $F=0.999$. The former is the value of the optimal state fidelity achieved in the previous part of our analysis. The latter sets a bound to the values of state fidelity achievable through our scheme, and serves as a guide to the eye. While an increasing phononic temperature is associated with a decrease of the state fidelity, the scheme appears to be robust, as far as fidelity is concerned. Such conclusions are validated and strengthened by a study of the phonon-number distribution in the conditional state of the mechanical oscillator for different values of its initial temperature, which is reported in Fig~\ref{fig:phostattemp}: Only for temperatures $\ge15$mK we see a significant contribution from states with $n\ge2$, thus ensuring the feasibility of the proposed protocol for temperatures that are within reach through standard passive cooling techniques.


\section{A steady state assessment of the protocol}
\label{sec:new}

We now briefly address the steady-state counterpart of the protocol discussed so far, highlighting relative performances and differences between the two schemes. We base our assessment on the proposal put forward by one of us in Ref.~\cite{paternostro2011engineering}, and consider a red-detuned working point with a lower mechanical effective mass ($m=5$pg), as it is the case in levitated optomechanics. We further assume that the system is addressed when stationary conditions are reached. We refer to Ref.~\cite{paternostro2011engineering} for details of the calculations, which are all along the lines of what has been presented here, and discuss directly the results of our analysis. We only stress that, in light with the discussions above, we use the initial temperature of the mechanical mode as the tuneable parameter for the evaluation of the performances of this steady-stets version of the SPNS-engineering protocol.

\begin{figure}[t]
\begin{center}  
{\bf(a)}\hskip3cm{\bf (b)}\\
\includegraphics[width=0.25\textwidth]{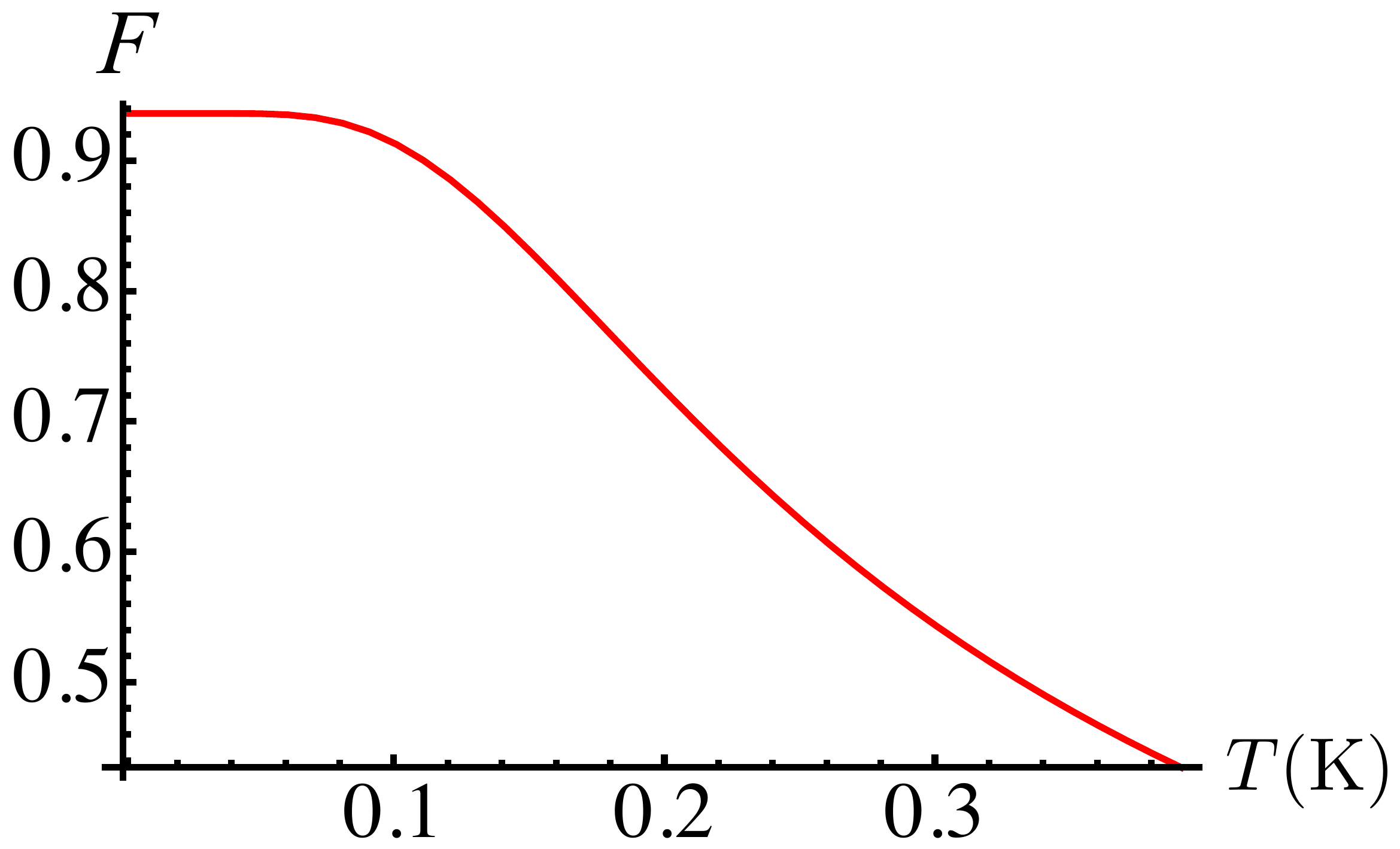}\includegraphics[width=0.25\textwidth]{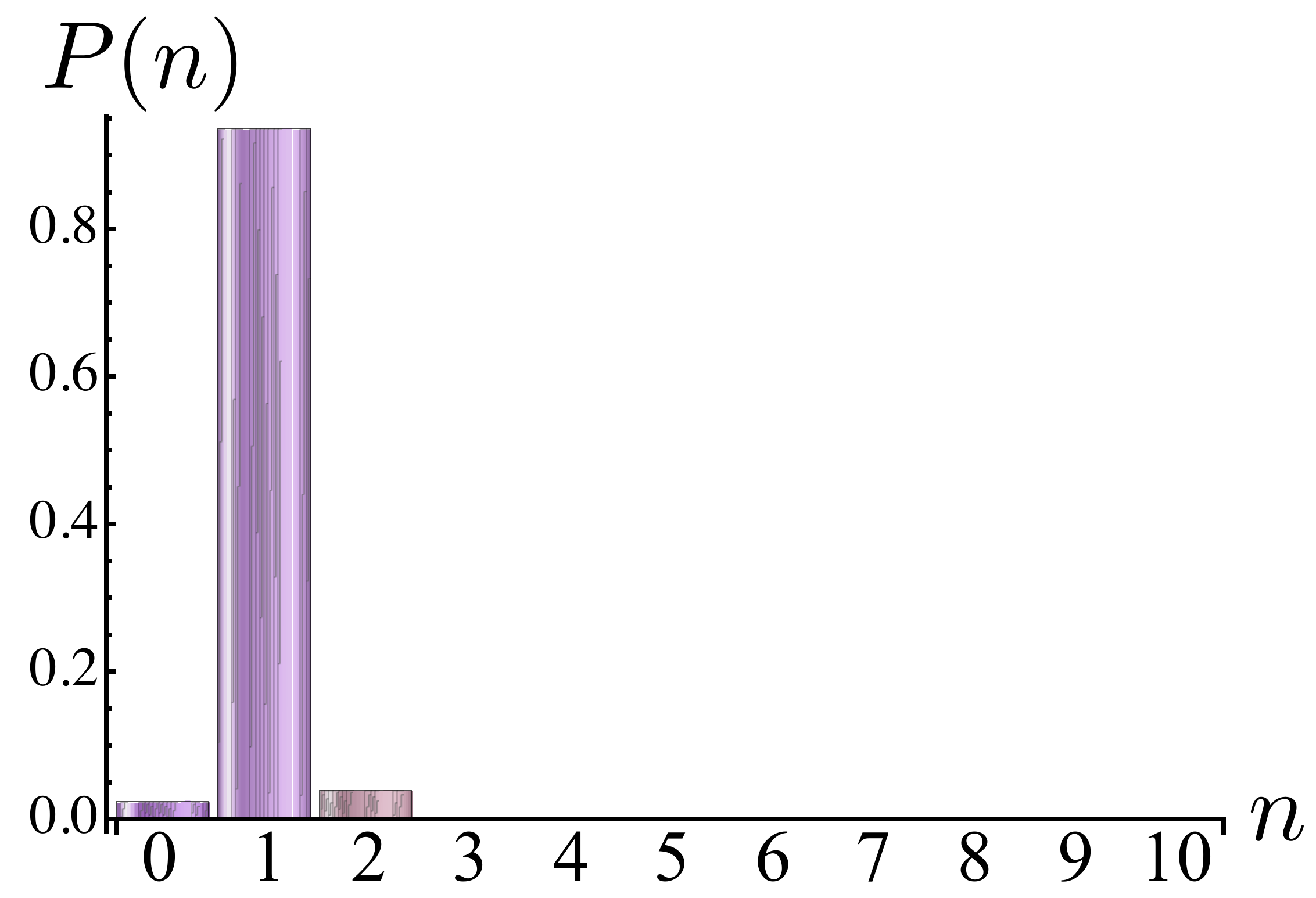}
\caption{(Color online) {\bf (a)} Fidelity between the conditional stationary mechanical state and a SPNS plotted against the initial mechanical temperature. {\bf (b)} The probability distribution of the conditional stationary state for $T=1$mK. Other parameters as in Ref.~\cite{paternostro2011engineering}.}
 \label{fidesteady}
\end{center}
\end{figure} 

Fig.~\ref{fidesteady} {\bf (a)} shows the behavior of the state fidelity with a SPNS as the temperature increases, showing the excellent performance of the protocol for an ample range of values of such parameter. Low temperatures perform significantly better, and we now concentrate on what is achieved by fixing $T=1$mK. However, we should remark the comparatively inferior performance, for the current choice of working point, with respect to the dynamical scheme illustrated above. Both features (the good similarity with a SPNS and the inferiority with respect to a time-gated photon subtraction) are consolidated in Fig.~\ref{fidesteady} {\bf (b)}, where we show the probability distribution of having $n$ excitations in the conditional mechanical state. At variance with the previous result, the contributions coming from the ground state and $|n=2\rangle$ are not insignificant, thus lowering the similarity with the desired target state. However, all the key features of a SPNS are retained by the conditional stationary mechanical state. 
More extensive analyses of the relative performance of the two approaches, including a detailed study of the differences in the types of quantum correlations shared by the optical and mechanical oscillators and the effects of additional subtractions steps on the form of the conditional mechanical state are left to further investigations. 

 
\section{Conclusions}\label{S6}
We have proposed a scheme for the engineering of SPNS of a massive mechanical mode based on a photon subtraction process. The effectiveness and robustness of the protocol has been assessed using relevant figures of merit and studying the effect of the temperature of the mechanical system, which is a key parameter in any optomechanical dynamics. We have argued that, at variance with a previously reported scheme for the achievement of non-classical mechanical states~\cite{paternostro2011engineering}, a dynamical approach with a time-gated photon-subtraction event, and a working point deep in the blue-detuning regime allow for the achievement of the best performances. Once engineered through the protocol illustrated herein, the mechanical state can be reconstructed using high-precision all-optical methods~\cite{mari2011directly,vanner2013cooling}. The proposal is thus fully within the reach of current state-of-the-art experiments in optomechanics, and paves the way to a novel approach towards the engineering of key non-Gaussian states of massive mechanical systems and their use for the quantum coherent communication. 

\acknowledgments
MP acknowledges support from the EU project TherMiQ, the John Templeton Foundation (grant number 43467), the Julian Schwinger Foundation (grant number JSF-14-7-0000), and the UK EPSRC (grant EP/M003019/1).

\section*{Appendix}

In this Appendix we give the explicit form of the functions entering the Wigner distribution of the conditional mechanical state given in Eq.~\eqref{wigner}. We have
\begin{widetext}
\begin{equation}\label{normalization}
\begin{aligned}
A^t_{0}&=\frac{{\cal N}}{\pi }\left(\frac{1}{m_{11}}\right)^{5/2} \sqrt{\frac{m_{11}}{4 m_{11} m_{22}-(m_{12}+m_{21})^2}},\quad {\cal N}=\frac{4}{f_{11}+f_{22}-2},\quad A^t_1=\frac{m_{11}^2} {4 m_{11}
   m_{22}-(m_{12}+m_{21})^2}{\cal P},\\
   {\cal P}&=-4 (c_{21}^2+c_{22}^2) m_{11}+4 (c_{11}
   c_{21}+c_{12} c_{22}) (m_{12}+m_{21})-4
   (c_{11}^2+c_{12}^2) m_{22}\\
   &-\left(f_{11}+f_{22}-2\right)
   [(m_{12}+m_{21})^2-4 m_{11} m_{22}],\\
B^t_{rr}&=\frac{16 m_{11}^2 [4 (c_{21}^2 + c_{22}^2) m_{11}^2-4 (c_{11} c_{21}+ c_{12} c_{22}) (m_{12}+m_{21})
   m_{11}+(c_{11}^2+c_{12}^2)
   (m_{12}+m_{21})^2]}{[4
   m_{11} m_{22}-(m_{12}+m_{21})^2]^2},\\
   B^t_{ri}&=\frac{32 m_{11}^2 \{2 (m_{12}+m_{21}) [(c_{21}^2+c_{22}^2) m_{11}+(c_{11}^2+c_{12}^2) m_{22}]-(c_{11} c_{21}+c_{12}
   c_{22}) [(m_{12}+m_{21})^2+4 m_{11} m_{22}]\}}{[4 m_{11} m_{22}-(m_{12}+m_{21})^2]^2},\\
   B^t_{ii}&=\frac{16 m_{11}^2 [(c_{21}^2+c_{22}^2) (m_{12}+m_{21})^2-4(c_{11} c_{21}+c_{12} c_{22})(m_{12}+m_{21}) m_{22} +4
   (c_{11}^2+c_{12}^2) m_{22}^2]}{[4 m_{11} m_{22}-(m_{12}+m_{21})^2]^2},\\
   C^t&=-\frac{8 \left(m_{22} \delta _i^2+\left(m_{12}+m_{21}\right) \delta _i \delta _r+m_{11} \delta _r^2\right)}{4 m_{11} m_{22}-(m_{12}+m_{21})^2}.
\end{aligned}
\end{equation}
\end{widetext}


\begin{thebibliography}{99}

\bibitem{aspelmeyer2014cavity} M. Aspelmeyer, T. J. Kippenberg, and F. Marquardt, {\it Cavity Optomechanics: Nano-and Micromechanical Resonators Interacting with Light} (Springer, 2014).
\bibitem{pikkalainen} J.-M. Pikkalainen, E. Damsk\"agg, M. Brandt, F. Massel, and M. A. Sillanp\"aa, Phys. Rev. Lett. {\bf 115}, 243601 (2015).
\bibitem{wollman} E. E. Wollman, C. U. Lei, A. J. Weinstein, J. Suh, A. Kronwald, F. Marquardt, A. A. Clerk, and K. C. Schwab, Science {\bf 28}, 952 (2015).
\bibitem{palomaki} T. A. Palomaki, J. W. Harlow, J. D. Teufel, R. W. Simmonds, and K. W. Lehnert, Nature {\bf 495}, 210 (2013).
\bibitem{riedinger} R. Riedinger, S. Hong, R. A. Norte, J. A. Slater, J. Shang, A. G. Krause, V. Anant, M. Aspelmeyer, and S. Gr\"oblacher, Nature {\bf 530}, 313 (2016).
\bibitem{schmidt} M. Schmidt, M. Ludwig, and F. Marquardt, New J. Phys. {\bf 14}, 125005 (2012).
\bibitem{houhou} O. Houhou, H. Aissaoui, and A. Ferraro, Phys. Rev. A {\bf 92}, 063843 (2015).
\bibitem{zurek2003decoherence} W. H. Zurek, quant-ph/0306072 (2003).
\bibitem{khalili2010preparing} F. Khalili, S. Danilishin, H. Miao, H. M\"uller-Ebhardt, H. Yang, and Y. Chen, Phys. Rev. Lett. {\bf 105}, 070403 (2010).
\bibitem{akram2010single} U. Akram, N. Kiesel, M. Aspelmeyer, and G. J. Milburn, New J. Phys. {\bf 12}, 083030 (2010).
\bibitem{paternostro2011engineering} M. Paternostro, Phys. Rev. Lett. {\bf 106}, 183601 (2011).
\bibitem{PhysRevA.83.013803} O. Romero-Isart, A. C. Pflanzer, M. L. Juan, R. Quidant, N. Kiesel, M. Aspelmeyer, and J. I. Cirac, Phys. Rev. A {\bf 83}, 013803 (2011).
\bibitem{tan2013deterministic} H. Tan, F. Bariani, G. Li, and P. Meystre, Phys. Rev. A {\bf 88}, 023817 (2013).
\bibitem{asjad2014reservoir} M. Asjad and D. Vitali, J. Phys. B: At., Mol. Opt. Phys. {\bf 47}, 5502 (2014).
\bibitem{PhysRevA.91.013842} W. Ge and M. S. Zubairy, Phys. Rev. A {\bf 91}, 013842 (2015).
\bibitem{akram2013entangled} U. Akram, W. P. Bowen, and G. J. Milburn, New J. Phys. {\bf 15}, 093007 (2013).
\bibitem{PhysRevA.89.053829} H. Tan, Phys. Rev. A {\bf 89}, 053829 (2014).
\bibitem{galland2014heralded} C. Galland, N. Sangouard, N. Piro, N. Gisin, and T. J. Kippenberg, Phys. Rev. Lett. {\bf 112}, 143602 (2014).
\bibitem{o2010quantum} A. D. O'Connell, M. Hofheinz, M. Ansmann, R. C. Bialczak, M. Lenander, E. Lucero, M. Neeley, D. Sank, H. Wang, M. Weides, et al., Nature {\bf 464}, 697 (2010).
\bibitem{aspelmeyer2013cavity} M. Aspelmeyer, T. J. Kippenberg, and F. Marquardt, Rev. Mod. Phys. {\bf 86}, 1391 (2014).
\bibitem{vitali2007optomechanical} D. Vitali, S. Gigan, A. Ferreira, H. B\"ohm, P. Tombesi, A. Guerreiro, V. Vedral, A. Zeilinger, and M. As- pelmeyer, Phys. Rev. Lett. {\bf 98}, 030405 (2007).
\bibitem{paternostro2006reconstructing} M. Paternostro, S. Gigan, M. Kim, F. Blaser, H. B\"ohm, and M. Aspelmeyer, New J. Phys. {\bf 8}, 107 (2006).
\bibitem{rogers2014hybrid} B. Rogers, N. Lo Gullo, G. De Chiara, G. M. Palma, and M. Paternostro, Quantum Measurements and Quantum Metrology {\bf 2}, 11 (2014).
\bibitem{ferreira2009quantum} A. Ferreira, arXiv:0911.2217 (2009).
\bibitem{hudson1974wigner} R. Hudson, Rep. Math. Phys. {\bf 6}, 249 (1974).
\bibitem{chan2011laser} J. Chan, T. M. Alegre, A. H. Safavi-Naeini, J. T. Hill, A. Krause, S. Gr\"oblacher, M. Aspelmeyer, and O. Painter, Nature {\bf 478}, 89 (2011).
\bibitem{chan2012optimized} J. Chan, A. H. Safavi-Naeini, J. T. Hill, S. Meenehan, and O. Painter, Appl. Phys. Lett. {\bf 101}, 081115 (2012). 
\bibitem{1742-6596-400-5-052024} D. H. Nguyen, A. Sidorenko, M. M\"uller, S. Paschen, A. Waard, and G. Frossati, J.  Phys.: Conf. Ser. {\bf 400} (2012).
\bibitem{ANDP:ANDP201400107} J. Zhang, T. Zhang, A. Xuereb, D. Vitali, and J. Li, Annalen der Physik {\bf 527}, 147 (2015).
\bibitem{hammerer2014nonclassical} K. Hammerer, C. Genes, D. Vitali, P. Tombesi, G. Milburn, C. Simon, and D. Bouwmeester, {\it Nonclassical states of light and mechanics} (Springer, 2014).
\bibitem{paternostro2008mechanism} M. Paternostro, J. Phys. B: At., Mol. Opt. Phys. {\bf 41}, 155503 (2008).
\bibitem{genes2008robust} C. Genes, A. Mari, P. Tombesi, and D. Vitali, Phys. Rev. A {\bf 78}, 032316 (2008).
\bibitem{horo} R. Horodecki, P. Horodecki, M. Horodecki, and K. Horodecki, Rev. Mod. Phys. {\bf 81}, 865 (2009). 
\bibitem{mari2011directly} A. Mari, K. Kieling, B. M. Nielsen, E. S. Polzik, and J. Eisert, Phys. Rev. Lett. {\bf 106}, 010403 (2011).
\bibitem{vanner2013cooling} M. Vanner, J. Hofer, G. Cole, and M. Aspelmeyer, Nature Comm. {\bf 4}, 2295 (2013).

\end{thebibliography}
\end{document}